\documentclass[a4paper,11pt]{article}
\usepackage[T1]{fontenc}
\usepackage{amsmath}
\usepackage{mathtools}
\usepackage{hyperref}
\usepackage{cite}
\usepackage{collref}
\usepackage{bbm}
\DeclareMathOperator{\dd}{d}

\newcommand{\SU}{\mathrm{SU}}
\newcommand{\U}{\mathrm{U}}
\newcommand{\Orth}{\mathrm{O}}
\newcommand{\Z}{\mathbbm{Z}}
\newcommand{\SW}{S_{\mbox{\tiny{W}}}}
\newcommand{\Tc}{T_{\mbox{\tiny{c}}}}
\newcommand{\nconf}{n_{\mbox{\tiny{conf}}}}
\newcommand{\betac}{\beta_{\mbox{\tiny{c}}}}
\newcommand{\SCFT}{S_{\mbox{\tiny{CFT}}}}
\newcommand{\ide}{\mathbbm{1}}
\newcommand{\Tr}{{\rm Tr\,}}
\newcommand{\Rmin}{R_{\mbox{\tiny{min}}}}
\newcommand{\Rmax}{R_{\mbox{\tiny{max}}}}
\newcommand{\redchisq}{\chi^2_{\tiny\mbox{red}}}

\begin{document}
\begin{titlepage}
\begin{flushright} 
DESY 19-073
\end{flushright}
\vskip0.5cm
\begin{center}
{\Large\bf Conformal perturbation theory confronts lattice results in the vicinity of a critical point}
\end{center}
\vskip1.3cm
\centerline{Michele~Caselle$^{1,2}$, Nicodemo~Magnoli$^{3}$, Alessandro~Nada$^{4}$, Marco~Panero$^{1}$, and Marcello~Scanavino$^{3}$}
\vskip1.5cm
\centerline{\sl $^{1}$ Department of Physics, University of Turin \& INFN, Turin}
\centerline{\sl Via Pietro Giuria 1, I-10125 Turin, Italy}
\vskip0.5cm
\centerline{\sl $^{2}$ Arnold-Regge Center, University of Turin}
\centerline{\sl Via Pietro Giuria 1, I-10125 Turin, Italy}
\vskip0.5cm
\centerline{\sl $^{3}$ Department of Physics, University of Genoa \& INFN, Genova}
\centerline{\sl Via Dodecaneso 33, I-16146, Genoa, Italy}
\vskip0.5cm
\centerline{\sl $^{4}$John von Neumann Institute for Computing, DESY}
\centerline{\sl Platanenallee 6, D-15738 Zeuthen, Germany}
\vskip0.5cm
\begin{center}
\href{mailto:caselle@to.infn.it}{{\tt caselle@to.infn.it}}, \href{mailto:nicodemo.magnoli@ge.infn.it}{{\tt nicodemo.magnoli@ge.infn.it}}, \href{mailto:alessandro.nada@desy.de}{{\tt alessandro.nada@desy.de}}, \href{mailto:marco.panero@unito.it}{{\tt marco.panero@unito.it}}, \href{mailto:marcello.scanavino@ge.infn.it}{{\tt marcello.scanavino@ge.infn.it}}
\end{center}
\vskip1.0cm
\begin{abstract}

We study the accuracy and predictive power of conformal perturbation theory by a comparison with lattice results in the neighborhood of the finite-temperature deconfinement transition of $\SU(2)$ Yang-Mills theory, assuming that the infrared properties of this non-Abelian gauge theory near criticality can be described by the Ising model. The results of this comparison show that conformal perturbation theory yields quantitatively accurate predictions in a broad temperature range. We discuss the implications of these findings for the description of the critical point (belonging to the same universality class) of another strongly coupled, non-supersymmetric non-Abelian gauge theory: the critical end-point in the phase diagram of QCD at finite temperature and finite quark chemical potential.
\end{abstract}

\end{titlepage}

\section{Introduction}
\label{sec:introduction}

Since the publication of a seminal article by Zamolodchikov~\cite{Zamolodchikov:1987ti}, conformal perturbation theory (CPT) has proved a powerful analytical tool to describe statistical-mechanics models and quantum field theories in the vicinity of a critical point. While its original application was limited to two-dimensional models, see for example refs.~\cite{Zamolodchikov:1990bk, Guida:1995kc, Guida:1996ux, Caselle:1999mg, Amoretti:2017aze}, recently it has also been extended to theories defined in three dimensions~\cite{Caselle:2015csa, Caselle:2016mww}, thanks to the recent developments in the calculation of Wilson coefficients using the conformal bootstrap approach~\cite{Komargodski:2016auf, Kos:2016ysd}.

In this work, we propose to use CPT to study the behavior of quantum chromodynamics (QCD) and other strongly coupled non-Abelian gauge theories near the critical points associated with a continuous phase transition in their phase diagram. The long-term goal of this approach is to derive theoretical predictions for the dynamics of strong interactions under the extreme conditions of temperature and density that are realized in heavy-ion collisions, for the values of center-of-mass energy and nuclear masses allowing one to probe the neighborhood of the critical end-point appearing at finite temperature $T$ and quark chemical potential $\mu$ in the QCD phase diagram. That critical point is  the end-point of the line of first-order transitions separating the hadronic from the deconfined phase (see refs.~\cite{BraunMunzinger:2008tz, Fukushima:2010bq, Luo:2017faz} for reviews). It is well known that such a line does not extend to the $\mu=0$ axis, where the deconfining and chiral-symmetry restoring transition is actually a crossover. Hence, the first-order transition line is believed to terminate at an end-point corresponding to a continuous phase transition, where the infrared behavior of the theory should be described by the critical exponents characteristic of an effective field theory compatible with the expected symmetry and dimension requirements, which is just the Ising model in three dimensions. In the past few years, the breakthrough based on the conformal bootstrap~\cite{Rattazzi:2008pe} allowed the analytical evaluation of critical exponents in the Ising model to an unprecedented level of precision~\cite{ElShowk:2012ht, El-Showk:2014dwa, Gliozzi:2013ysa, Gliozzi:2014jsa, Kos:2014bka, Simmons-Duffin:2015qma, Kos:2016ysd}, in some cases superior to the corresponding Monte~Carlo estimates by $2$ orders of magnitude.

The experimental search for the QCD critical end-point, proposed $21$ years ago~\cite{Stephanov:1998dy}, remains a very active line of research~\cite{Lacey:2006bc, Aggarwal:2010wy, Adamczyk:2013dal, Lacey:2014wqa, Gazdzicki:2015ska, Adare:2015aqk, Adamczyk:2017wsl, Yin:2018ejt}. Meanwhile, however, an \emph{ab initio} derivation of the existence and location of this critical end-point is still missing~\cite{Stephanov:2004wx}: this is mainly due to the fact that the tool of choice for theoretical studies of strong interactions in the regime probed in heavy-ion collisions, namely numerical calculations in the lattice regularization~\cite{Wilson:1974sk} is obstructed by a computational sign problem when the quark chemical potential is finite~\cite{deForcrand:2010ys, Gattringer:2016kco}. As a consequence, complementary theoretical approaches could provide valuable information on the region of the phase diagram in the neighborhood of the critical end-point. We argue that conformal perturbation theory could allow one to study the physics of strongly coupled QCD matter at values of temperature and net baryonic density lying along the ``trajectories'' (in the phase diagram) scanned in experiments, as long as such trajectories pass sufficiently close to the critical point.

It is important to understand how well conformal perturbation theory works at a quantitative level, i.e. how large is the range of parameters of the underlying microscopic theory (in this case, QCD), for which the resulting low-energy physics can be approximated well by the associated conformal model (in this case, the Ising model in three dimensions) at or near criticality.

To this purpose, in this work we present a detailed comparison of theoretical predictions from conformal perturbation theory, with those derived numerically using lattice simulations. We do this for $\SU(2)$ Yang-Mills theory, a strongly coupled non-Abelian gauge theory in four spacetime dimensions exhibiting a continuous phase transition at a finite deconfinement temperature $\Tc$, which is in the same universality class~\cite{Svetitsky:1982gs} as the one associated with the critical end-point of QCD, namely the one of the Ising model in three dimensions. In contrast to the critical end-point of QCD, however, the critical behavior at the deconfinement transition in finite-temperature $\SU(2)$ Yang-Mills theory can be studied numerically to very high precision, making this theory a useful proxy with which to test the quantitative accuracy of conformal-perturbation-theory predictions.

In particular, we focus our comparison on the two-point correlation function of Wilson lines winding around in the Euclidean-time direction: an important observable in lattice gauge theory, which, at $T=0$, can be directly linked to the potential $V(r)$ of a pair of static color sources a distance $r$ apart from each other, and, as a consequence, to the spectrum of heavy-quark bound states~\cite{Bali:2000gf, Brambilla:2010cs} (for a classic study of this quantity in $\SU(2)$ Yang-Mills theory, see ref.~\cite{Bali:1994de}).

Presently, much analytical information is known about the behavior of the potential derived from this correlator in non-supersymmetric non-Abelian gauge theories: at short distances, asymptotic freedom implies that its dominant contribution is a Coulomb term, arising from one-gluon exchange, and the separation between the momentum and mass scales allows one to organize the different terms appearing in perturbative calculations in a systematic way~\cite{Brambilla:1999xf}. Conversely, the long-distance physics is dominated by non-perturbative features, resulting in a linearly rising potential $V(r)$ at asymptotically large $r$: assuming the formation of a confining flux tube, with energy per unit length $\sigma$ (the ``string tension''), it is then possible to show that its dominant excitations in the infrared limit are described by massless bosonic oscillations in the transverse direction, that yield a characteristic $1/r$ correction to $V(r)$~\cite{Luscher:1980fr, Luscher:1980ac}. This picture can be described by a low-energy effective theory, associated with the spontaneous breaking of translational and rotational symmetries in the presence of a confining string~\cite{Aharony:2013ipa}: the requirement of a non-linear realization of Lorentz-Poincar\'e invariance poses very tight constraints on the terms of this effective theory, making it highly predictive~\cite{Aharony:2013ipa,Billo:2012da}: for a recent, comprehensive review, including a discussion of lattice results, see ref.~\cite{Brandt:2016xsp}.

The situation in finite-temperature QCD is more subtle, due to the presence of the additional energy scale defined by the temperature $T$~\cite{Brambilla:2004jw, Brambilla:2008cx, Brambilla:2010vq, Brambilla:2010xn, Brambilla:2011sg, Brambilla:2013dpa} and to the non-trivial r\^ole of infrared divergences~\cite{Linde:1980ts, Gross:1980br}, with screening and damping effects. The modern, proper definition of the potential between heavy color sources has a real and an imaginary part, which can be reconstructed from a spectral-function analysis of thermal Wilson loops~\cite{Rothkopf:2011db, Burnier:2014ssa}. Nevertheless, the Euclidean correlator of Polyakov lines (and $-T$ times its logarithm, which, for simplicity, we still denote as $V$) still encodes interesting information about the thermal behavior of the theory. In the confining phase, the long-distance properties of this correlator can be accurately modeled assuming that the flux tube joining the color sources oscillates with a Euclidean action proportional to the area it spans, i.e. that in the infrared limit the dynamics of the theory is described by a low-energy effective action equal to the Nambu-Got{\={o}} action~\cite{Nambu:1974zg, Goto:1971ce}. At intermediate quark-antiquark distances $r$, however, deviations from the ideal picture of a Nambu-Got{\={o}} string do show up, as well as corrections induced by the finite temperature~\cite{Pisarski:1982cn}. In fact, it has been known for a long time that the Polyakov-loop operator captures much of the dynamics of Yang-Mills theory close to the deconfinement temperature (see, for example, ref.~\cite{Ogilvie:1983ss}).

The purpose of this work is to study numerically, by Monte~Carlo simulations on the lattice, the two-point correlation function of Polyakov lines in $\SU(2)$ Yang-Mills theory, at temperatures in the vicinity of its second-order deconfinement transition, and to compare the simulation results with analytical calculations in conformal perturbation theory. As will be discussed below, the main findings of this work are:
\begin{itemize}
\item The results obtained from conformal perturbation theory are in very good agreement with those from lattice simulations in a rather wide temperature interval; i.e. conformal perturbation theory provides reliable predictions in a rather large neighborhood of the conformal point.
\item Conformal perturbation theory predicts the Polyakov-loop correlator to be described by an operator-product expansion (OPE) with \emph{different} coefficients above and below the critical temperature $\Tc$; the ratio of these coefficients is fixed by the universality class of the conformal model, and can be predicted in conformal field theory. The numerical values of the coefficients extracted from our lattice simulations of the $\SU(2)$ Yang-Mills theory above and below $\Tc$ are such, that their ratio agrees with the value predicted by conformal theory for the Ising universality class.
\end{itemize}

This article is organized as follows. In section~\ref{sec:CPT} we briefly review the main features of conformal perturbation theory, focusing on the formulas relevant for this work. Section~\ref{sec:su2results} presents the setup and results of our lattice simulation of $\SU(2)$ Yang-Mills theory, as well as their comparison with the predictions of conformal perturbation theory. Finally, section~\ref{sec:conclusions} includes a detailed discussion about the applicability of conformal perturbation theory to the study of the critical end-point of QCD and some concluding remarks. Preliminary results of this work have been reported in ref.~\cite{Caselle:2018jdu}.

\section{Conformal perturbation theory}
\label{sec:CPT}

Conformal perturbation theory is a mathematical tool to work out an expansion for the short-distance behavior of correlation functions of quantum field theories, in the vicinity of a conformally invariant critical point. In particular, following ref.~\cite{Guida:1995kc}, CPT can be seen as a way to derive the coefficients of the Wilson operator-product expansion~\cite{Wilson:1969zs} that is induced, when a conformally invariant theory is perturbed by a relevant operator. The method deals with short-distance divergences in a standard fashion, and is self-consistent in the long-distance limit, where it yields finite results: this is a clear advantage over more conventional expansions, say, in powers of the mass perturbing the conformal theory, which are often plagued by infrared divergences.

Technically, the key aspect of CPT expansions is that, by construction, they clearly separate the non-perturbative features of the theory (i.e. the vacuum expectation values of different operators) from those that can be computed perturbatively (i.e. the Wilson coefficients). Another important feature of CPT is that it only requires the knowledge of limited information characterizing the theory~\cite{Guida:1995kc, Caselle:1999mg}: this includes universal (like the critical indices) as well as non-universal data, (like critical amplitudes of one-point functions, which can be obtained using  off-critical methods, such as strong- or weak-coupling expansions, or numerical simulations).

The calculation of off-critical correlators by means of CPT has greatly benefited from the recent progress in the determination of universal quantities by the conformal-bootstrap method (see ref.~\cite{Poland:2018epd} for a recent review): in particular, accurate predictions have been worked out for the perturbations of conformal models in the universality class of the three-dimensional Ising model~\cite{Caselle:2015csa, Caselle:2016mww}. 

CPT predictions for this universality class can be \emph{directly} tested against lattice results in the vicinity of the critical point associated with the finite-temperature deconfining phase transition in purely gluonic Yang-Mills theory with an $\SU(2)$ gauge group. In this case the correspondence between the degrees of freedom of the gauge theory and those of the spin model is clear~\cite{Svetitsky:1982gs}: $n$-point correlation functions of thermal Wilson lines (or Polyakov loops) are mapped to $n$-point correlators of spin degrees of freedom. In particular, in this work we are interested in the behavior of the two-point correlator of Polyakov loops in the gauge theory, which, denoting the spin degrees of freedom by $\sigma$, is mapped to the $\langle \sigma (r) \sigma (0) \rangle$ correlator in the Ising model.

The CPT analysis for the three-dimensional Ising model is straightforward. This model is characterized by two relevant operators, namely the energy density $\epsilon$ and the magnetization (the one-point correlation function of the spin $\sigma$); the dimensions of these operators have been recently computed and are $\Delta_\epsilon=1.412625(10)$ and $\Delta_\sigma=0.5181489(10)$~\cite{Komargodski:2016auf, Kos:2016ysd}. The action of the perturbed model is defined as
\begin{equation}
\label{action}
S = \SCFT + t \int \dd ^3 x \, \epsilon(x),
\end{equation}
where $\SCFT$ denotes the action at the critical point, and the parameter $t$ is related to the deviation from the critical temperature of the model. For the non-critical theory at finite $t$, the behavior of the two-point correlation function of operators $\mathcal{O}_i$ and $\mathcal{O}_j$ at short separation $r$ can be expressed in terms of the Wilson coefficients $C_{ijk}$ appearing in the expansion:
\begin{equation}
\label{Wilson_expansion}
\langle \mathcal{O}_i (r) \mathcal{O}_j (0) \rangle_t = \sum_k C_{ijk}(r,t)\langle \mathcal{O}_k \rangle_t.
\end{equation}
The Wilson coefficients can be expanded in Taylor series in $t$,
\begin{equation}
\label{Taylor_for_Cijk}
C_{ijk}(r,t) = \sum_{n=0}^\infty \frac{t^n}{n!} \frac{\partial^n C_{ijk}}{\partial t^n}.
\end{equation}
The partial derivatives appearing on the right-hand side of this equation are not divergent at large $r$. Defining $\Delta_t = 3-\Delta_\epsilon$ and writing the one-point correlation functions for the energy density and magnetization at finite $t$ as
\begin{equation}
\langle \epsilon \rangle_t = A^{\pm} |t|^{\frac{\Delta_\epsilon}{\Delta_t}},  \qquad \langle \sigma \rangle_t = B_{\sigma} |t|^{\frac{\Delta_\sigma}{\Delta_t}}
\end{equation}
(see also ref.~\cite{Pelissetto:2000ek}), the leading terms in the conformal perturbative expansion of the $\langle \sigma(r) \sigma(0) \rangle_t$ correlator are
\begin{equation}
\label{correlator_expansion}
\langle \sigma (r) \sigma (0) \rangle _t = C_{\sigma\sigma\ide}(0,r) + 
C_{\sigma\sigma\epsilon}(0,r) A^{\pm} |t|^{\frac{\Delta_\epsilon}{\Delta_t}} +
t \partial_{t} C_{\sigma\sigma\ide}(0,r) + \dots ,
\end{equation}
where $\ide$ is the identity operator, while $C_{\sigma\sigma\ide}(0,r)=r^{-2 \Delta_\sigma}$ and $C_{\sigma\sigma\epsilon}(0,r)=c_{\sigma\sigma\epsilon} r^{\Delta_\epsilon-2\Delta_\sigma}$ denote the Wilson coefficients evaluated at the critical point. Finally, the partial derivative of $C_{\sigma\sigma\ide}$ appearing in eq.~(\ref{correlator_expansion}) can be written as
\begin{equation}
\label{partial_derivative_of_Css1_wrt_t}
\partial_t C_{\sigma \sigma \ide} (0,r) = k_{\sigma\sigma\ide} r^{\Delta_t - 2 \Delta_\sigma} c_{\sigma\sigma\epsilon}.
\end{equation}
with $k_{\sigma\sigma\ide} \simeq - 62.5336$: as discussed in detail in the part of text between eqs.~(6) and (7) of ref.~\cite{Caselle:2016mww}, $\partial_t C_{\sigma \sigma \ide} (0,r)$ can be computed by a Mellin transform, and reduced to a combination of Euler integrals of the second kind, which are functions of $\Delta_\epsilon$. Note that eq.~(\ref{correlator_expansion}) defines the non-connected correlation function.

Introducing the combination
\begin{equation}
\label{combination}
s=r^{\Delta_t}t,
\end{equation}
eq.~(\ref{correlator_expansion}) can be rewritten as
\begin{equation}
\label{correlator_expansion_2}
r^{2\Delta_\sigma} \langle \sigma(r) \sigma(0) \rangle_t = 1 + 
c_{\sigma\sigma\epsilon} A^{\pm} |s|^{\frac{\Delta_\epsilon}{\Delta_t}} + k_{\sigma\sigma\ide}c_{\sigma\sigma\epsilon}s + \dots 
\end{equation}

In order to compare this analytical prediction from CPT with the numerical results from Monte~Carlo simulations of $\SU(2)$ Yang-Mills theory, we have to fix some non-universal quantities. These include the following:
\begin{itemize}
\item The normalization of the Polyakov loop, i.e. a proportionality factor relating the $\sigma$ spin expectation value in the Ising model, and the Polyakov loop $P$ of $\SU(2)$ Yang-Mills theory, evaluated on the lattice. It should be noted that the latter quantity is a \emph{bare} one, which would tend to zero in the continuum limit~\cite{Dotsenko:1979wb, Mykkanen:2012ri}. As a consequence, the proportionality factor relating $\sigma$ and $P$ is a function of the lattice spacing of the Yang-Mills theory $a$, or, equivalently, of the Wilson parameter $\beta=4/g^2$, where $g$ denotes the bare lattice coupling. In this work, we fix this normalization by matching the two-point correlation function of Polyakov loops at the critical point to the corresponding spin-spin correlator in the critical Ising model.
\item Identifying $r$ with the spatial separation $R$ between the Polyakov lines, we reabsorb all non-universal factors into the definition of the perturbation coefficient $t$ in the spin model. This quantity is related to the perturbing parameter of the $\SU(2)$ lattice gauge theory, which is $\beta-\betac(N_t)$, where $\betac(N_t)$ is the value of the Wilson parameter (or of the bare gauge coupling) corresponding to a lattice spacing $a$ such that $aN_t$ is the inverse of the critical deconfinement temperature in natural units. Note that the $\beta-\betac(N_t)$ difference controls the deviation of the temperature from its critical value. To fix the non-universal relation between $t$ and $\beta-\betac(N_t)$, we take advantage of the universality of the last term on the right-hand side of eq.~(\ref{correlator_expansion_2}), by fitting our results for the correlator as a function of $r$, and using the result to fix the relation between $t$ and $\beta-\betac(N_t)$.
\item The amplitudes $A^{\pm}$ can be determined using the second term in the expansion above. The numerical value of these amplitudes is one of the non-trivial results of our analysis; the $A^+/A^-$ ratio is universal, and this expectation provides a useful check of the self-consistency and robustness of the whole analysis.
\end{itemize}

\section{Numerical results for $\SU(2)$ Yang-Mills theory}
\label{sec:su2results}

In order to test the predictions discussed in the previous section, we studied the behavior of the Polyakov loop correlators in the vicinity of the deconfinement transition of the $3+1$-dimensional $\SU(2)$ Yang-Mills theory. In the following subsection~\ref{subsec:setup}, we define the setup of our lattice regularization of this theory; then, we present our numerical results, comparing them with CPT predictions in subsection~\ref{subsec:comparison_with_CPT}, and discussing their uncertainties in subsection~\ref{subsec:systematic_uncertainties}.

\subsection{Setup of the lattice calculation}
\label{subsec:setup}

We regularize the theory on a finite hypercubic lattice of spacing $a$ and sizes $aN_t$ in the $\hat{0}$ (``Euclidean-time'') direction and $L=aN_s$ in the three other (``spatial'') directions, labeled as $\hat{1}$, $\hat{2}$, and $\hat{3}$. We always take $aN_s \gg aN_t$. The fundamental degrees of freedom of the lattice theory are matrices $U_\mu(x)$, taking values in the defining representation of the $\SU(2)$ group, and associated with parallel transporters between neighboring sites $x$ and $x+a\hat{\mu}$. Periodic boundary conditions are assumed in all directions. The dynamics of the theory is defined by the Wilson action~\cite{Wilson:1974sk}
\begin{equation}
\SW = -\frac{2}{g^2} \sum_{x} \sum_{0 \le \mu < \nu \le 3} \Tr U_{\mu\nu} (x)
\end{equation}
where $U_{\mu\nu}(x)=U_\mu(x)U_\nu\left(x+a\hat{\mu}\right)U_\mu^\dagger\left(x+a\hat{\nu}\right)U_\nu^\dagger(x)$ is a plaquette having the site $x$ as a corner and lying in the oriented $(\mu,\nu)$ plane, and $g^2$ is the squared bare coupling; we also introduce the parameter $\beta=4/g^2$.

The temperature $T$ is related to the extent of the shortest compact size of the lattice as $a N_t=1/T$: as a consequence, $T$ can be varied by changing $N_t$, $a$, or both. The physical value of the lattice spacing $a$ is set non-perturbatively, as discussed in ref.~\cite{Caselle:2015tza}, and is controlled by the parameter $\beta$. We express our results in terms of the deconfinement temperature $\Tc$, using the value for the ratio of $\Tc$ over the square root of the zero-temperature string tension $\Tc/\sqrt{\sigma}=0.7091(36)$, which was reported in ref.~\cite{Lucini:2003zr}.

The Polyakov loop at a spatial coordinate $\vec{x}$ is defined as the trace of the closed Wilson line in the $\hat{0}$ direction:
\begin{equation}
P\left(\vec{x}\right) = \frac{1}{2} \Tr \prod_{t=0}^{N_t} U_0 \left(ta,\vec{x}\right).
\end{equation}
The two-point correlation function of Polyakov loops is then defined as
\begin{equation}
\label{def_G}
G(R) = \left\langle \frac{1}{N_s^3} \sum_{\vec{x}} \ P\left(\vec{x}\right) P\left(\vec{x}+R \hat{k}\right) \right\rangle ,
\end{equation}
where $\hat{k}$ denotes one of the three ``spatial'' directions, the sum is over all spatial coordinates $\vec{x}$, while the $\langle \dots \rangle$ average is taken over all values of all of the $U_\mu(x)$ variables, with a measure that is proportional to the product of the Haar measures of all $U_\mu(x)$ matrices and to $\exp(-\SW)$, and normalized in such a way that the expectation value of the identity operator is $1$.

We remark that, like eq.~(\ref{correlator_expansion}), eq.~(\ref{def_G}) defines the \emph{non-connected} (i.e. full) correlator, in which the square of the average value of the Polyakov loop (which, in the thermodynamic limit, is non-zero in the deconfined phase) is \emph{not} subtracted. The reason for this choice is that we are going to compare the lattice results for this correlator with the analogous correlator in conformal perturbation theory, where the correlation function of interest is the non-connected one~\cite{Zamolodchikov:1990bk, Guida:1995kc, Guida:1996ux, Caselle:1999mg, Amoretti:2017aze}. The fact that the CPT formalism deals with the non-connected correlators (i.e. does not encode any information on long-wavelength physics, including a possibly non-vanishing vacuum expectation value of the field) is unsurprising, given that it is ultimately formulated in terms of a particular type of operator-product expansion, which is expected to capture the behavior of physical correlators at short distances only. Accordingly, as will be discussed below (see also the values reported in table~\ref{tab:results}), we will restrict our fit ranges to distances shorter than the characteristic correlation length of the theory at that temperature.

One may wonder whether it might be possible to carry out a meaningful fit to CPT using the connected correlator, too. The answer is no: the reason is that, as discussed in section~\ref{sec:CPT}, the non-trivial information from CPT is expressed in terms of a function of a non-trivial combination of the variables that describe the temperature and the distance, see eqs.~(\ref{combination}) and~(\ref{correlator_expansion_2}), whereas the quantity that is subtracted from the full correlator to obtain the connected one, i.e. the square of the average value of the Polyakov loop, is, by definition, $R$ independent, but \emph{temperature dependent}. 

Finally, it is also worth noting that, by contrast, the fact that the lattice correlator defined in eq.~(\ref{def_G}) is a \emph{bare} one does not hinder the possibility of a comparison with CPT predictions, thanks to the fact that the Polyakov loop undergoes a purely multiplicative renormalization~\cite{Dotsenko:1979wb, Mykkanen:2012ri}.

% we emphasize that, in agreement with the definitions of the relevant formulas in the CPT literature that we use, the two-point function to be used is not the connected one, and, accordingly, the definition of the lattice correlator already appearing in the previous version of our manuscript is correct. We emphasized this aspect in the text after equation (5), pointing out that, on the one hand, the OPE formulas that are used in the conformal perturbation theory approach are not expected to capture the infrared dynamics, and that, on the other hand, the CPT formulas depend on a precise particular combination of the variables describing the distance and the temperature, so that one could not simply compare them with a lattice two-point correlation function in which, for the deconfined phase, a temperature-dependent, but r-independent, term is subtracted. Finally, we also stressed that, accordingly, the fits to CPT that we carried out were always restricted to distances shorter than the corresponding correlation length at that temperature (as the comparison of the values in the third and fourth columns of table 3 shows), hence the imperfect agreement observed in the deconfined phase (and discussed in the last part of subsection 3.2) cannot be directly interpreted as due to a missing constant term.

In the confining phase (that is, for $T<\Tc$) we fit our numerical results for $G(R)$ to the functional form
\begin{equation}
\label{G_fit}
G(R) = A \left\{ \frac{\exp\left( -R/\xi \right)}{R} + \frac{\exp\left[ - (L-R)/\xi \right]}{L-R} \right\},
\end{equation}
with $\xi$ (which is the largest correlation length of the theory) and $A$ (which is an overall amplitude, with no direct physical meaning) as fitting parameters. Note that the second summand on the right-hand side of eq.~(\ref{G_fit}) accounts for the effect of the closest periodic copy of the Polyakov line on the hypertorus; we neglect the effect of other periodic copies (at distances $L$, $\sqrt{L^2+R^2}$, $\sqrt{R^2+2L(L-R)}$, \dots) as well as corrections due to higher-energy states.

As expected in the presence of a continuous phase transition, $\xi \to \infty$ for $T \to \Tc$. More precisely, in the proximity of the critical point, $\xi$ diverges like $[(T-\Tc)/\Tc]^{-\nu}$ (which defines the hyperscaling exponent $\nu$), with two different \emph{amplitudes}, respectively denoted as $\xi_{0^+}$ and as $\xi_{0^-}$, for $T>\Tc$ and $T<\Tc$. While these amplitudes are not universal, their ratio is, and for the universality class of the Ising model that ratio was evaluated to be $\xi_{0^+}/\xi_{0^-}=1.95(2)$ in refs.~\cite{Caselle:1997dc, Caselle:1997hs}. This allows one to obtain an estimate of the typical correlation length also in the deconfined phase (at least for temperatures not very far from $\Tc$). The characteristic correlation length estimated this way provides one with an upper bound for the range of distances over which the numerical results from lattice simulations of the $\SU(2)$ Yang-Mills theory can be compared with the analytical predictions from conformal perturbation theory: in all of the fits that we carried out, we always restricted our comparison of the $G(R)$ correlator with CPT predictions to distances not larger than a maximum Polyakov-loop separation $\Rmax$, with $\Rmax\ll\xi$. At the same time, the shortest $R$ distances probed in the fits are always larger than a few units of the lattice spacing $a$. The double constraint $ a \ll R \ll \xi $ enforces the hierarchy of scales making a sensible comparison between lattice results and CPT predictions possible.

Table~\ref{tab:lattice_setup} summarizes the parameters of the Monte~Carlo simulations carried out in the present work.

\begin{table}[!htb]
\centering
\begin{tabular}{|c|c|c|c|c|}
\hline
 $N_t \times N_s^3$ & $\beta$    & $T/T_c$ & $\nconf$         & $\xi/a$     \\
\hline                             
\hline                             
  $8\times 80^3$    & $2.48479$  & $0.90$  & $8\times 10^4$   & $9.24(3)$   \\
                    & $2.50311$  & $0.96$  & $8\times 10^4$   & $23.3(2)$   \\
                    & $2.50598$  & $0.97$  & $8\times 10^4$   & $43.3(4)$   \\
                    & $2.51165$  & $1$     & $8\times 10^4$   &             \\
                    & $2.52295$  & $1.02$  & $8\times 10^4$   & $\sim 85$   \\
                    & $2.52567$  & $1.05$  & $8\times 10^4$   & $\sim 45$   \\
                    & $2.54189$  & $1.10$  & $8\times 10^4$   & $\sim18$    \\
\hline                             
\hline                             
 $10\times 80^3$    & $2.55$     & $0.90$  & $10^5$           & $11.72(8)$  \\
                    & $2.569$    & $0.96$  & $10^5$           & $29.4(3)$   \\
                    & $2.572$    & $0.97$  & $10^5$           & $42.9(4)$   \\
                    & $2.58101$  & $1$     & $8\times 10^4$   &             \\
                    & $2.58984$  & $1.02$  & $1.6\times 10^5$ & $\sim 85$   \\
                    & $2.59271$  & $1.05$  & $1.6\times 10^5$ & $\sim 55$   \\
                    & $2.61$     & $1.10$  & $1.6\times 10^5$ & $\sim23$    \\
\hline
\hline
 $12\times 96^3$    & $2.60573$  & $0.90$  & $8\times10^4$    & $12.89(15)$ \\
                    & $2.626$    & $0.96$  & $8\times10^4$    & $34.8(4)$   \\
                    & $2.62923$  & $0.97$  & $8\times10^4$    & $41.3(3)$   \\
                    & $2.63896$  & $1$     & $8\times10^4$    & \\
                    & $2.64558$  & $1.02$  & $1.6\times10^5$  & $\sim 81$   \\
                    & $2.65541$  & $1.05$  & $1.6\times10^5$  & $\sim 65$   \\
                    & $2.67085$  & $1.10$  & $1.6\times10^5$  & $ \sim 25$  \\
\hline
\end{tabular}
\caption{Information about the parameters of our lattice calculations for $\SU(2)$ Yang-Mills theory. The first two columns show the lattice sizes in units of the lattice spacing $a$ and the parameter $\beta=4/g^2$, while in the third we display the temperature in units of $\Tc$ and in the fourth the statistics for the Polyakov-loop correlators. Finally, in the last column we present our estimates for the correlation length $\xi$, in units of $a$.
\label{tab:lattice_setup}}
\end{table}

\subsection{Comparison with CPT predictions}
\label{subsec:comparison_with_CPT}

We analyzed our lattice results for the Polyakov-loop correlators in $\SU(2)$ Yang-Mills theory as follows.
\begin{enumerate}
\item First, we fixed the normalization constant for the Polyakov loops by matching the value of $G(R)$ at the critical temperature $T=\Tc$ to the corresponding quantity in the Ising model at criticality, i.e. the spin-spin correlator.
\item Then, we fitted the numerical value of the correlator to eq.~(\ref{correlator_expansion_2}), as a function of $R$, keeping the coefficients of the second and third term on the right-hand side of that equation as the parameters to be fitted. 
\item Finally, we used our best estimates for these coefficients to fix the remaining quantities, and studied how they depend on the temperature $T$.
\end{enumerate}

\begin{figure}
\centering
\includegraphics[scale=1.0]{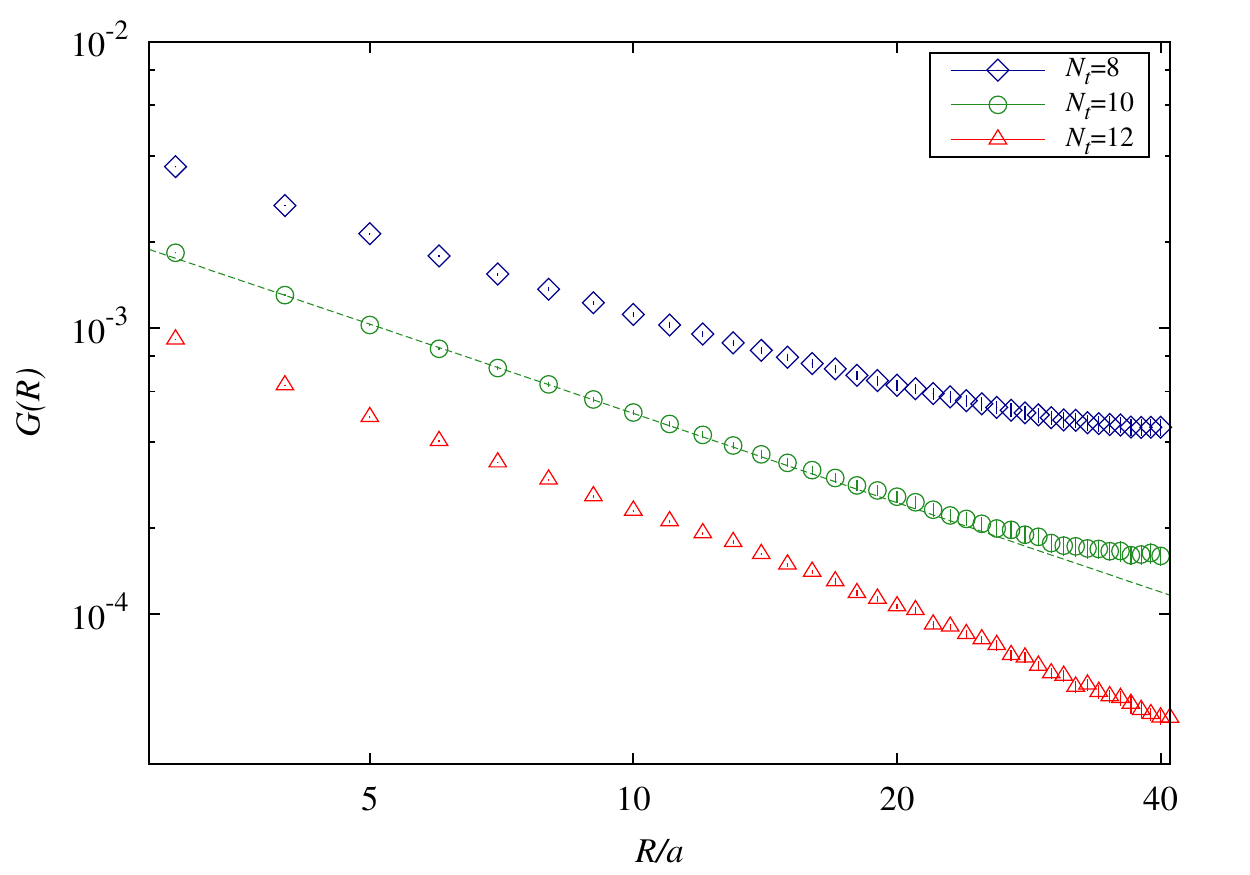}
\caption{\label{fig:correlators_at_Tc} Logarithmic plot of our results for the two-point correlation function of Polyakov loops $G(R)$ in the $\SU(2)$ gauge theory at the deconfinement temperature, as a function of the distance $R$.}
\end{figure}

For the first of these steps, fig.~\ref{fig:correlators_at_Tc} shows an example of our results for the $G(R)$ correlator at the critical temperature: the lattice data confirm the expected power-law behavior (revealing itself as a straight line in the plot with logarithmic axes), and the presence of significant finite-size effects for the points at the largest values of $R$.

Thus, we fit the correlator at criticality to the form
\begin{equation}
\label{critical_correlator}
 G(R)=\frac{C_P^2}{R^{2\Delta_\sigma}}
\end{equation}
for different ranges of values of the Polyakov-loop separation $\Rmin \le R \le \Rmax$. In order to avoid excessive contamination from lattice discretization artifacts (on the gauge-theory side) and/or from other charge-conjugation-odd operators (in the comparison with the conformal field theory), we set $\Rmin=4a$, and fitted the data for different values of $\Rmax$. An example of the results of this analysis (from a lattice with $N_t=10$ at $T=\Tc$) is shown in tab.~\ref{tab:fit_at_Tc}. As expected, the data at the largest values of $R$ (close to $L/2$) are affected by non-negligible contamination due to the periodic copies of the lattice. Combining the results from the fits with $\Rmax=12a$ and $\Rmax=20a$, we take $C_P^2=0.00547(2)$ as our final estimate for the critical-correlator fit at this value of $N_t$.

\begin{table}
\begin{center}
\begin{tabular}{|c|c|c|c|}
\hline
$\Rmin/a$ & $\Rmax/a$ & $C_{P}^2$      & $\redchisq$ \\
\hline
$4$       & $12$      & $0.005463(12)$ & $0.25$      \\
\hline
$4$       & $20$      & $0.005477(15)$ & $0.40$      \\
\hline
$4$       & $30$      & $0.005492(19)$ & $0.66$      \\
\hline
$4$       & $40$      & $0.005513(31)$ & $1.96$      \\
\hline
\end{tabular}
\caption{Results of our fits of the Polyakov-loop correlator $G(R)$ at $N_t=10$, $N_s=80$ and $\beta=2.58101$, corresponding to $T=\Tc$, to eq.~(\ref{critical_correlator}). The results for $C_{P}^2$, shown in the third column, are obtained from fits for $\Rmin \le R \le \Rmax$ (first two columns); the values of the reduced $\chi^2$ are listed in the last column.}
\label{tab:fit_at_Tc}
\end{center}
\end{table}

\begin{figure}
 \centering
 \includegraphics[scale=1.0]{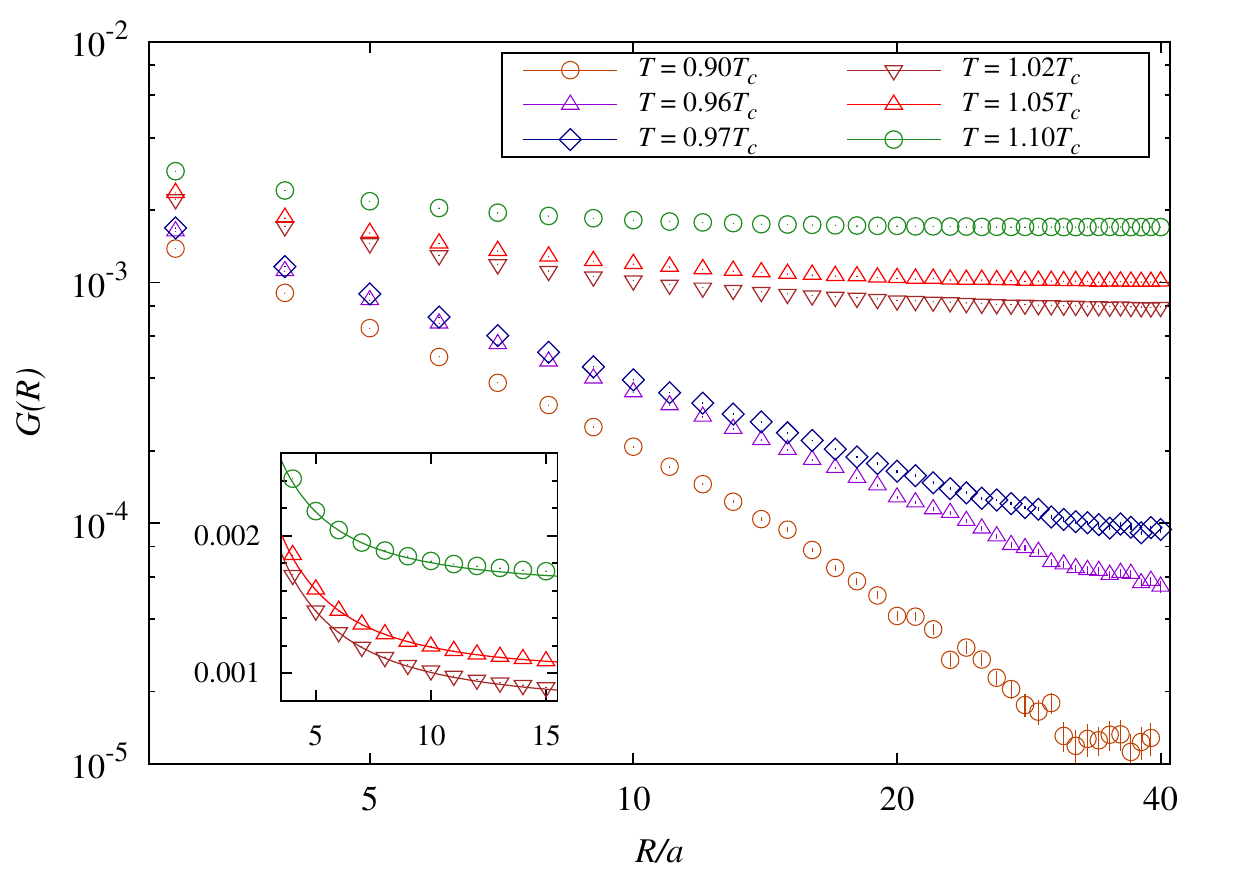}
 \caption{Example of results for the Polyakov-loop two-point correlation function (in logarithmic scale), plotted against the spatial separation $R$ (in linear scale), for different temperatures $T$ off $\Tc$. These results were obtained from simulations on lattices with $N_t=10$ lattice spacings in the Euclidean-time direction. The inset (in which both axes are in a linear scale) shows an enlargement of the region where the correlators at $T>\Tc$ were fitted to the CPT prediction, eq.~(\ref{correlator_at_T_different_from_Tc}), as discussed in the text.}
 \label{fig:offcritical_nt10}
\end{figure}

For the analysis of the Polyakov loop correlators $G(R)$ at $T \neq \Tc$ we fitted the results of the correlator to the functional form
\begin{equation}
\label{correlator_at_T_different_from_Tc}
G(R) = \frac{C^2_P}{R^{2\Delta_\sigma}} \left( 1 + c_1 R^{\Delta_\epsilon} + c_2 R^{\Delta_t} \right),
\end{equation}
where the exponents $\Delta_\sigma$, $\Delta_\epsilon$ and $\Delta_t$ are those discussed in section~\ref{sec:CPT}, while $c_1$ and $c_2$ are the free parameters. The results of this analysis are reported in table~\ref{tab:results} and shown in fig.~\ref{fig:offcritical_nt10}, where the inset shows a zoom onto the smaller range of distances, where the results at $T>\Tc$ are fitted.

\begin{table}
\begin{center}
\begin{tabular}{|c|c|c|c|c|c|}
\hline
$\beta$   & $T/T_c$  & $\Rmax/a$ & $\xi/a$    & $c_1$       & $c_2$         \\
\hline
$2.55$    & $0.90$   & $[7-8]$   & $11.72(8)$ & $-0.169(1)$ & $0.099(1)$    \\
$2.569$   & $0.96$   & $[11-14]$ & $29.4(3)$  & $-0.067(2)$ & $0.037(1)$    \\
$2.572$   & $0.97$   & $[12-21]$ & $42.9(4)$  & $-0.048(3)$ & $0.026(2)$    \\
$2.58984$ & $1.02$   & $[18-25]$ & $\sim 85$  & $0.067(2)$  & $-0.019(1)$   \\
$2.59271$ & $1.05$   & $[13-19]$ & $\sim 55$  & $0.091(2)$  & $-0.0256(15)$ \\
$2.61$    & $1.10$   & $[8-9]$   & $\sim 23$  & $0.221(3)$  & $-0.081(3)$   \\
\hline
\end{tabular}
\caption{Example of results of the fits of the correlator $G(R)$ to eq.~(\ref{correlator_at_T_different_from_Tc}), obtained from simulations on lattices with $N_t=10$ at different values of $\beta=4/g^2$ (first column), corresponding to the temperatures reported in the second column, in the range $4a \le R \le \Rmax$, and for the values of $\Rmax$ shown in the third column. In the fourth column, we display our estimates for the correlation lengths in units of the lattice spacing, while the fitted parameters $c_1$ and $c_2$ are listed in the fifth and in the sixth column, respectively. Note that, as discussed in the text, at each temperature, the largest distances at which the correlators are fitted (shown in the third column) are always chosen to be shorter, or much shorter, than the corresponding correlation lengths reported in the fourth column.}
\label{tab:results}
\end{center}
\end{table}

Next, we investigated the relation between the perturbing parameter $t$ and the difference $\beta-\betac(N_t)$ using the following relation:
\begin{equation}
\label{c2_relation}
c_2 = k_{\sigma\sigma\ide} \cdot C^{\epsilon}_{\sigma\sigma} \cdot t.
\end{equation}
Using the values for $k_{\sigma\sigma\ide} \simeq -62.5336$ and for $C^{\epsilon}_{\sigma \sigma} = 1.0518537(41)$ reported in ref.~\cite{Caselle:2016mww} and in refs.~\cite{Komargodski:2016auf, Kos:2016ysd}, respectively, the analysis of the data set corresponding to $N_t=10$ yields the values for $t$ reported in tab.~\ref{tab:t_latt}.

\begin{table}
\begin{center}
\begin{tabular}{|c|c|c|c|}
\hline
$\beta$   & $T/T_c$ & $t$             & $\Delta\beta$  \\
\hline
$2.55$    & $0.90$  & $0.001505(15)$  & $-0.01550$     \\
$2.569$   & $0.96$  & $0.000563(15)$  & $-0.00600$     \\
$2.572$   & $0.97$  & $0.000395(30)$  & $-0.00450$     \\
$2.58984$ & $1.02$  & $-0.000284(18)$ & $0.00440$      \\
$2.59271$ & $1.05$  & $-0.000389(23)$ & $0.00585$      \\
$2.61$    & $1.10$  & $-0.001231(45)$ & $0.01450$      \\
\hline
\end{tabular}
\caption{Results for the perturbing parameter $t$ and for $\Delta\beta$, obtained from eq.~(\ref{c2_relation}), at different values of the temperature in the proximity of the deconfining transition. The analysis is based on a set of data obtained from simulations on a lattice with $N_t=10$ lattice spacings in the Euclidean-time direction.}
\label{tab:t_latt}
\end{center}
\end{table}

The table also reports the values of $\Delta\beta=[\beta -\beta_c(N_t)]/2$, which for the $\SU(2)$ lattice gauge theory is the perturbing parameter with respect to the plaquette operator $\sum_{x} \sum_{0 \le \mu < \nu \le 3} \Tr U_{\mu\nu} (x)$. As expected, $t$ has a negative sign in the deconfined phase, where the center symmetry is broken, and a positive sign in the confining, $\Z_2$-symmetric phase. The magnitude of these values of $t$ is similar to those studied in ref.~\cite{Caselle:2016mww}, for which conformal perturbation theory was found to give reliable predictions, which leads us to expect that this should also be the case here.

We note that, interestingly, $t$ is almost exactly proportional to $\Delta\beta$: this means that, in the neighborhood of the critical temperature, the energy operator of the three-dimensional Ising model encodes the dynamics of the Euclidean-action density operator of the four-dimensional $\SU(2)$ gauge theory in a quantitatively accurate way.

By fixing $t$, it is possible to derive the values of the amplitudes $A^+$ (in the confining phase, i.e. for $T<\Tc$) and $A^-$ (at $T>\Tc$) from the relation
\begin{equation}
A^\pm=\mp \frac{c_1}{C^{\epsilon}_{\sigma\sigma}}|t|^{-\frac{\Delta_\epsilon }{\Delta_t }}.
\end{equation}

The determination of these amplitudes allows for a non-trivial test of the validity of this CPT analysis: in particular, $A^+$ should be constant and negative in the confining phase, while in the deconfined phase the amplitude $A^-$ should be constant and positive. Furthermore, the $A^+/A^-$ ratio should be universal and is predicted to be $A^+/A^-=-0.536(2)$~\cite{Hasenbusch:2010ua}.

\begin{table}
\begin{center}
\begin{tabular}{|c|c|c|c|}
\hline
$\beta$   & $T/\Tc$ & $A^+$      & $A^-$   \\
\hline
$2.55$    & $0.90$  & $-52.6(6)$ &         \\
$2.569$   & $0.96$  & $-50(2)$   &         \\
$2.572$   & $0.97$  & $-49(5)$   &         \\
$2.58984$ & $1.02$  &            & $91(6)$ \\
$2.59271$ & $1.05$  &            & $94(5)$ \\
$2.61$    & $1.10$  &            & $82(3)$ \\
\hline
\end{tabular}
\caption{Results for the amplitude $A^+$ in the confining phase (first three rows) and for $A^-$ in the deconfining phase (last three rows), for the values of inverse coupling $\beta$ and of the temperature $T$ reported in the first two columns, as discussed in the text.}
\label{tab:amplitudes}
\end{center}
\end{table}

Our results for $A^\pm$ are reported in table~\ref{tab:amplitudes}: in the confining phase, the values for $A^+$ are indeed compatible with a constant (that we estimate as $A^+=-50(2)$), while in the deconfined phase the values are slightly less stable, with a quantitatively significant deviation at the largest temperature. This may indicate that the largest temperature that we investigated is close to the limit where our leading-order conformal-perturbation-theory analysis breaks down. We remark that, as our fits to CPT were limited to spatial separations shorter (or much shorter) than the correlation length at that temperature, this slight instability of the fits in the deconfined phase is not simply interpretable as an effect caused by the fact that our fits do not include an $R$-independent term. Rather, it may be a numerical artifact induced by the fact that, in the deconfined phase, as the temperature is increased to values larger and larger than $\Tc$, the correlation length decreases, and, as a consequence, the fitting range in $R$ (whose upper limit is bound to be shorter than the correlation length) reduces to only a few points. 

We estimate the amplitude in the deconfined phase to be $A^-=89(6)$. Accordingly, we obtain the numerical value $A^+/A^-=-0.56(4)$, which is compatible with the one computed in ref.~\cite{Hasenbusch:2010ua}.

\subsection{Systematic uncertainties}
\label{subsec:systematic_uncertainties}

We conclude this section with a few comments on the uncertainties involved in our analysis.

In order to test the impact of finite-volume effects, we repeated a subset of our calculations also on lattices at a larger value of $N_s$ (the extent of the system in each of the spatial directions), namely $N_s=96$. In all cases, we found that the results obtained from simulations with $N_s=96$ were compatible with those at $N_s=80$ within their uncertainties.\footnote{We also observed that, on smaller lattices with $N_s=64$, some deviations start to be visible, at least for temperatures sufficiently close to $\Tc$.}

While in this work we have not carried out a systematic study of the continuum limit, we remark that, in addition to the results discussed here (from the analysis of data obtained at $N_t=10$), we also repeated the analysis for those at $N_t=8$ and at $N_t=12$, finding the same qualitative picture.

One may wonder, why conformal perturbation theory describes the dynamics of this gauge theory so well. The main reason is that, for the underlying conformal theory that is involved in this case (i.e. the one describing the Ising model in three dimensions at criticality), the conformal weights of the terms included in the expansion in eq.~(\ref{correlator_expansion_2}) and the subleading terms, that are neglected, are separated by a finite (and sufficiently large) gap. More precisely, the terms appearing in the parentheses on the right-hand side of eq.~(\ref{correlator_expansion_2}), besides the constant, have exponents (for $t$) which are approximately equal to $0.6$ and $1.2$, while the first subleading correction, that is neglected in eq.~(\ref{correlator_expansion_2}), scales at least quadratically in $t$, see ref.~\cite[eq.~(5)]{Caselle:2015csa}. In turn, this feature of the conformal spectrum for the three-dimensional Ising universality class is due to the intrinsic simplicity of the operator content of the model. 

Our results show that conformal perturbation theory works well in a wide neighborhood of the critical point, and at least for temperatures down to $T/T_c=0.90$. While in principle it would be possible to test the CPT predictions at even lower temperatures, this would require significantly finer lattices (and, as a consequence, computationally much more demanding simulations), in order to keep the physical correlation length well separated from the lattice spacing.

\section{Proposed extension for the study of the QCD critical point and concluding remarks}
\label{sec:conclusions}

In this work we presented a numerical test of conformal perturbation theory as a tool to predict the behavior of strongly interacting gauge theories in the proximity of a critical point.

Specifically, we focused on $\SU(2)$ Yang-Mills theory at finite temperature $T$: this theory, which can be studied to high precision by numerical calculations on the lattice, has a second-order deconfinement phase transition at a finite temperature $\Tc$, which is in the universality class of the three-dimensional Ising model. Accordingly, correlation functions in the gauge theory at criticality are mapped to those in the spin model, and universality arguments imply a set of interesting predictions for the behavior of the gauge theory at $T=\Tc$. Conformal perturbation theory extends these predictions from the critical \emph{point} to a whole \emph{finite neighborhood}: as we discussed in detail in section~\ref{sec:su2results}, the physical correlation functions of the strongly coupled gauge theory near the critical point can be successfully described by means of the corresponding truncated conformal perturbation theory expansions, such as eq.~(\ref{correlator_expansion_2}). The approximation involved in the truncation is robust, as long as the terms that are neglected are sufficiently suppressed. For the universality class of the Ising model in three dimensions, this is indeed the case, and conformal perturbation theory successfully predicts the behavior of the gauge theory in a large interval of temperatures.\footnote{Note that this approach does \emph{not} rely on the existence of a whole line of critical points. For a discussion of the predictivity of conformal field theory for models with one or more critical lines, see for example ref.~\cite{Caselle:2019khe} and the bibliography therein.}

This successful test of conformal perturbation theory opens up the possibility of interesting generalizations. Of particular relevance is the one for the critical end-point in the phase diagram of QCD, which, if it exists, is expected to be in the universality class of the Ising model in three dimensions.

On the theoretical side, it should be emphasized that the origin of the critical point in the QCD phase diagram is totally different from the one in the purely gluonic $\SU(2)$ Yang-Mills theory, with $N=2$ color charges, that we considered here. As we discussed above, the latter is the critical point associated with a phase transition taking place at finite temperature and at zero quark chemical potential $\mu$. The absence of dynamical fields in representations of non-vanishing $N$-ality implies that, at the classical level, the theory has an exact $\Z_2$ global symmetry, associated with the center of the gauge group. The thermal deconfinement phase transition is then interpreted as the breakdown of this symmetry, taking place at a finite temperature $\Tc$, which separates the confining, $\Z_2$-symmetric phase at $T<\Tc$, from a deconfined phase at $T>\Tc$, in which this center symmetry is dynamically broken. This transition is known to be of second order~\cite{Engels:1990vr, Fingberg:1992ju, Engels:1994xj} and, in agreement with expectations from universality and from renormalization-group arguments~\cite{Svetitsky:1982gs}, its critical exponents are consistent with those of the three-dimensional spin system with a $\Z_2$ global symmetry group, i.e. the Ising model~\cite{condmat0012164}.

By contrast, for $\SU(N)$ gauge theories with $N>2$ color charges the thermal deconfinement transition is discontinuous~\cite{Lucini:2002ku, Lucini:2003zr, Lucini:2005vg, Panero:2009tv, Datta:2009jn, Datta:2010sq, Lucini:2012wq, Lucini:2012gg}. In QCD, which includes dynamical quark fields in the fundamental representation of the gauge group, the center symmetry is  absent even at the classical level, being explicitly broken by the Dirac operator. In turn, the fact that the \emph{up} and \emph{down} quark flavors have very small masses implies that classically the QCD Lagrangian has an approximate $\U(2) \times \U(2)$ symmetry, in which one can identify different components: in addition to ``vector'' $\U(1)$ (baryon-number conservation) and $\SU(2)$ (isospin symmetry) terms, which are preserved at the quantum level and manifest in the hadronic spectrum at $T=0$, the ``axial'' $\U(1)$ is anomalous~\cite{'tHooft:1976up, 'tHooft:1976fv, Fujikawa:1979ay}, while the existence of a non-vanishing quark condensate $\langle \bar{\psi} \psi \rangle$ leads to dynamical symmetry breaking and to the interpretation of the three pions as Nambu-Goldstone bosons. The fact that the masses of the \emph{up} and \emph{down} quarks are small but not exactly zero, implies that chiral symmetry is not an actual symmetry of the QCD Lagrangian either. As a consequence, even though the quark condensate vanishes at a finite temperature and for $\mu=0$, this change of state from the hadronic phase to a deconfined, chirally symmetric phase is not a transition, but rather a crossover~\cite{Aoki:2006br, Bazavov:2011nk}. When a non-vanishing \emph{net} baryon-antibaryon number is allowed, through a non-vanishing chemical potential $\mu$, the QCD phase diagram is expected to reveal interesting novel phases~\cite{Alford:1998mk, Alford:2006wn, Stephanov:2007fk}, and model calculations suggest the presence of a line of first-order transitions separating the hadronic phase from the quark-gluon plasma phase. If that line exists, then it should turn into the crossover band at a critical end-point\footnote{If the phase diagram is extended by the inclusion of a third axis, to study the dependence on the mass of the two light quark flavors $m$, then for $m=0$ at zero and small $\mu$ one expects the two phases to be separated by a second-order phase transition in the universality class of the $\Orth(4)$ spin model in three dimensions~\cite{Pisarski:1983ms}, although the restoration of the axial $\U(1)$ symmetry could change this scenario~\cite{Pelissetto:2013hqa}.} at finite values of $T$ and $\mu$, at which the long-distance physics is, again, expected to be described in terms of a conformal field theory in the universality class of the liquid-gas phase transition, i.e. in the universality class of the three-dimensional Ising model~\cite{Halasz:1998qr, Berges:1998rc} (see also ref.~\cite{Hatta:2002sj}). Recent works in this direction include refs.~\cite{Antoniou:2018ico, Parotto:2018pwx, Pradeep:2019ccv, Martinez:2019bsn}.

The discussion above clarifies that, although the symmetries of $\SU(2)$ Yang-Mills theory at finite temperature (and vanishing chemical potential) and those of QCD with dynamical quarks of physical masses at finite temperature and finite chemical potential are remarkably different, their critical behavior at the deconfinement phase transition and at the QCD critical end point are remarkably described by the same universality class, i.e. their static, long-range properties are expected to be those characteristic of the Ising model in three dimensions.

We already mentioned that the critical point of the QCD phase diagram is inaccessible to theoretical first-principle methods: it lies in a region which is far from the domain of applicability of perturbative computations, and out of reach for lattice calculations, due to the presence of the sign problem~\cite{deForcrand:2010ys, Gattringer:2016kco}. The nature of the latter is more profound than what one could na{\"{\i}}vely imagine for a purely computational problem~\cite{Troyer:2004ge}, making it unlikely that this fundamental obstruction to explore the QCD phase diagram could be overcome by a sheer increase in computing power, at least for classical computers.\footnote{In principle, new developments in quantum computing might have the potential to tackle the sign problem~\cite{Cloet:2019wre, Lamm:2019bik}.}

The location of the critical point in the QCD phase diagram, however, can be studied experimentally. From this point of view, the analysis of net-proton and net-kaon multiplicity distributions observed in heavy-ion collision experiments remains a key tool: for example, in refs.~\cite{Adamczyk:2013dal, Adamczyk:2017wsl} the STAR Collaboration presented results on the dependence of the mean, standard deviation, skewness and kurtosis of such distributions on the beam energy and centrality, and, in fact, a primary goal of the BES program at RHIC is to look for the first-order transition line and the QCD critical point~\cite{Abelev:2009bw}.

If the location of the QCD critical point were determined to sufficient precision, then it would be possible to formulate theoretical predictions in its neighborhood using conformal perturbation theory, much like we did in the present work. This will require an identification of the ``directions'' in the QCD phase diagram that correspond to different types of perturbations of the three-dimensional Ising model, and this will then allow one to derive a whole class of phenomenological predictions (in particular for particle distributions, correlators, and pion interferometry) of direct relevance for experiments. Working out analytical predictions for the strong nuclear interaction in a highly non-perturbative regime by means of conformal perturbation theory remains an exciting prospect for the future.

\subsection*{Acknowledgments} 
The lattice calculations were run on the MARCONI supercomputer of the Consorzio Interuniversitario per il Calcolo Automatico dell'Italia Nord Orientale (CINECA).

\bibliography{paper}

\begin{thebibliography}{100}

\bibitem{Zamolodchikov:1987ti}
A.~B. Zamolodchikov,
\newblock Sov. J. Nucl. Phys. {\bf 46}, 1090 (1987),
\newblock [Yad. Fiz. {\bf 46}, 1819 (1987)].

\bibitem{Zamolodchikov:1990bk}
Al.~B. Zamolodchikov,
\newblock Nucl. Phys. {\bf B348}, 619 (1991).

\bibitem{Guida:1995kc}
R.~Guida and N.~Magnoli,
\newblock Nucl. Phys. {\bf B471}, 361 (1996), hep-th/9511209.

\bibitem{Guida:1996ux}
R.~Guida and N.~Magnoli,
\newblock Nucl. Phys. {\bf B483}, 563 (1997), hep-th/9606072.

\bibitem{Caselle:1999mg}
M.~Caselle, P.~Grinza, and N.~Magnoli,
\newblock Nucl. Phys. {\bf B579}, 635 (2000), hep-th/9909065.

\bibitem{Amoretti:2017aze}
A.~Amoretti and N.~Magnoli,
\newblock Phys. Rev. {\bf D96}, 045016 (2017), 1705.03502.

\bibitem{Caselle:2015csa}
M.~Caselle, G.~Costagliola, and N.~Magnoli,
\newblock Phys. Rev. {\bf D91}, 061901 (2015), 1501.04065.

\bibitem{Caselle:2016mww}
M.~Caselle, G.~Costagliola, and N.~Magnoli,
\newblock Phys. Rev. {\bf D94}, 026005 (2016), 1605.05133.

\bibitem{Komargodski:2016auf}
Z.~Komargodski and D.~Simmons-Duffin,
\newblock J. Phys. {\bf A50}, 154001 (2017), 1603.04444.

\bibitem{Kos:2016ysd}
F.~Kos, D.~Poland, D.~Simmons-Duffin, and A.~Vichi,
\newblock JHEP {\bf 08}, 036 (2016), 1603.04436.

\bibitem{BraunMunzinger:2008tz}
P.~Braun-Munzinger and J.~Wambach,
\newblock Rev. Mod. Phys. {\bf 81}, 1031 (2009), 0801.4256.

\bibitem{Fukushima:2010bq}
K.~Fukushima and T.~Hatsuda,
\newblock Rept. Prog. Phys. {\bf 74}, 014001 (2011), 1005.4814.

\bibitem{Luo:2017faz}
X.~Luo and N.~Xu,
\newblock Nucl. Sci. Tech. {\bf 28}, 112 (2017), 1701.02105.

\bibitem{Rattazzi:2008pe}
R.~Rattazzi, V.~S. Rychkov, E.~Tonni, and A.~Vichi,
\newblock JHEP {\bf 12}, 031 (2008), 0807.0004.

\bibitem{ElShowk:2012ht}
S.~El-Showk {\em et~al.},
\newblock Phys. Rev. {\bf D86}, 025022 (2012), 1203.6064.

\bibitem{El-Showk:2014dwa}
S.~El-Showk {\em et~al.},
\newblock J. Stat. Phys. {\bf 157}, 869 (2014), 1403.4545.

\bibitem{Gliozzi:2013ysa}
F.~Gliozzi,
\newblock Phys. Rev. Lett. {\bf 111}, 161602 (2013), 1307.3111.

\bibitem{Gliozzi:2014jsa}
F.~Gliozzi and A.~Rago,
\newblock JHEP {\bf 10}, 042 (2014), 1403.6003.

\bibitem{Kos:2014bka}
F.~Kos, D.~Poland, and D.~Simmons-Duffin,
\newblock JHEP {\bf 11}, 109 (2014), 1406.4858.

\bibitem{Simmons-Duffin:2015qma}
D.~Simmons-Duffin,
\newblock JHEP {\bf 06}, 174 (2015), 1502.02033.

\bibitem{Stephanov:1998dy}
M.~A. Stephanov, K.~Rajagopal, and E.~V. Shuryak,
\newblock Phys. Rev. Lett. {\bf 81}, 4816 (1998), hep-ph/9806219.

\bibitem{Lacey:2006bc}
R.~A. Lacey {\em et~al.},
\newblock Phys. Rev. Lett. {\bf 98}, 092301 (2007), nucl-ex/0609025.

\bibitem{Aggarwal:2010wy}
STAR, M.~M. Aggarwal {\em et~al.},
\newblock Phys. Rev. Lett. {\bf 105}, 022302 (2010), 1004.4959.

\bibitem{Adamczyk:2013dal}
STAR, L.~Adamczyk {\em et~al.},
\newblock Phys. Rev. Lett. {\bf 112}, 032302 (2014), 1309.5681.

\bibitem{Lacey:2014wqa}
R.~A. Lacey,
\newblock Phys. Rev. Lett. {\bf 114}, 142301 (2015), 1411.7931.

\bibitem{Gazdzicki:2015ska}
M.~Gazdzicki and P.~Seyboth,
\newblock Acta Phys. Polon. {\bf B47}, 1201 (2016), 1506.08141.

\bibitem{Adare:2015aqk}
PHENIX, A.~Adare {\em et~al.},
\newblock Phys. Rev. {\bf C93}, 011901 (2016), 1506.07834.

\bibitem{Adamczyk:2017wsl}
STAR, L.~Adamczyk {\em et~al.},
\newblock Phys. Lett. {\bf B785}, 551 (2018), 1709.00773.

\bibitem{Yin:2018ejt}
Y.~Yin,
\newblock (2018), 1811.06519.

\bibitem{Stephanov:2004wx}
M.~A. Stephanov,
\newblock Prog. Theor. Phys. Suppl. {\bf 153}, 139 (2004), hep-ph/0402115,
\newblock [Int. J. Mod. Phys. A {\bf 20}, 4387(2005)].

\bibitem{Wilson:1974sk}
K.~G. Wilson,
\newblock Phys. Rev. {\bf D10}, 2445 (1974).

\bibitem{deForcrand:2010ys}
P.~de~Forcrand,
\newblock PoS {\bf Lattice 2009}, 010 (2009), 1005.0539.

\bibitem{Gattringer:2016kco}
C.~Gattringer and K.~Langfeld,
\newblock Int. J. Mod. Phys. {\bf A31}, 1643007 (2016), 1603.09517.

\bibitem{Svetitsky:1982gs}
B.~Svetitsky and L.~G. Yaffe,
\newblock Nucl. Phys. {\bf B210}, 423 (1982).

\bibitem{Bali:2000gf}
G.~S. Bali,
\newblock Phys. Rept. {\bf 343}, 1 (2001), hep-ph/0001312.

\bibitem{Brambilla:2010cs}
N.~Brambilla {\em et~al.},
\newblock Eur. Phys. J. {\bf C71}, 1534 (2011), 1010.5827.

\bibitem{Bali:1994de}
G.~Bali, K.~Schilling, and C.~Schlichter,
\newblock Phys. Rev. {\bf D51}, 5165 (1995), hep-lat/9409005.

\bibitem{Brambilla:1999xf}
N.~Brambilla, A.~Pineda, J.~Soto, and A.~Vairo,
\newblock Nucl. Phys. {\bf B566}, 275 (2000), hep-ph/9907240.

\bibitem{Luscher:1980fr}
M.~L{\"u}scher, K.~Symanzik, and P.~Weisz,
\newblock Nucl. Phys. {\bf B173}, 365 (1980).

\bibitem{Luscher:1980ac}
M.~L{\"u}scher,
\newblock Nucl. Phys. {\bf B180}, 317 (1981).

\bibitem{Aharony:2013ipa}
O.~Aharony and Z.~Komargodski,
\newblock JHEP {\bf 1305}, 118 (2013), 1302.6257.

\bibitem{Billo:2012da}
M.~Bill{\'o}, M.~Caselle, F.~Gliozzi, M.~Meineri, and R.~Pellegrini,
\newblock JHEP {\bf 1205}, 130 (2012), 1202.1984.

\bibitem{Brandt:2016xsp}
B.~B. Brandt and M.~Meineri,
\newblock Int. J. Mod. Phys. {\bf A31}, 1643001 (2016), 1603.06969.

\bibitem{Brambilla:2004jw}
N.~Brambilla, A.~Pineda, J.~Soto, and A.~Vairo,
\newblock Rev. Mod. Phys. {\bf 77}, 1423 (2005), hep-ph/0410047.

\bibitem{Brambilla:2008cx}
N.~Brambilla, J.~Ghiglieri, A.~Vairo, and P.~Petreczky,
\newblock Phys. Rev. {\bf D78}, 014017 (2008), 0804.0993.

\bibitem{Brambilla:2010vq}
N.~Brambilla, M.~{\'A}. Escobedo, J.~Ghiglieri, J.~Soto, and A.~Vairo,
\newblock JHEP {\bf 09}, 038 (2010), 1007.4156.

\bibitem{Brambilla:2010xn}
N.~Brambilla, J.~Ghiglieri, P.~Petreczky, and A.~Vairo,
\newblock Phys. Rev. {\bf D82}, 074019 (2010), 1007.5172.

\bibitem{Brambilla:2011sg}
N.~Brambilla, M.~{\'A}. Escobedo, J.~Ghiglieri, and A.~Vairo,
\newblock JHEP {\bf 12}, 116 (2011), 1109.5826.

\bibitem{Brambilla:2013dpa}
N.~Brambilla, M.~{\'A}. Escobedo, J.~Ghiglieri, and A.~Vairo,
\newblock JHEP {\bf 05}, 130 (2013), 1303.6097.

\bibitem{Linde:1980ts}
A.~D. Linde,
\newblock Phys. Lett. {\bf B96}, 289 (1980).

\bibitem{Gross:1980br}
D.~J. Gross, R.~D. Pisarski, and L.~G. Yaffe,
\newblock Rev. Mod. Phys. {\bf 53}, 43 (1981).

\bibitem{Rothkopf:2011db}
A.~Rothkopf, T.~Hatsuda, and S.~Sasaki,
\newblock Phys. Rev. Lett. {\bf 108}, 162001 (2012), 1108.1579.

\bibitem{Burnier:2014ssa}
Y.~Burnier, O.~Kaczmarek, and A.~Rothkopf,
\newblock Phys. Rev. Lett. {\bf 114}, 082001 (2015), 1410.2546.

\bibitem{Nambu:1974zg}
Y.~Nambu,
\newblock Phys. Rev. {\bf D10}, 4262 (1974).

\bibitem{Goto:1971ce}
T.~Got{\={o}},
\newblock Prog. Theor. Phys. {\bf 46}, 1560 (1971).

\bibitem{Pisarski:1982cn}
R.~D. Pisarski and O.~Alvarez,
\newblock Phys. Rev. {\bf D26}, 3735 (1982).

\bibitem{Ogilvie:1983ss}
M.~Ogilvie,
\newblock Phys. Rev. Lett. {\bf 52}, 1369 (1984).

\bibitem{Caselle:2018jdu}
M.~Caselle, N.~Magnoli, A.~Nada, M.~Panero, and M.~Scanavino,
\newblock PoS {\bf Confinement 2018}, 042 (2019), 1811.09208.

\bibitem{Wilson:1969zs}
K.~G. Wilson,
\newblock Phys. Rev. {\bf 179}, 1499 (1969).

\bibitem{Poland:2018epd}
D.~Poland, S.~Rychkov, and A.~Vichi,
\newblock Rev. Mod. Phys. {\bf 91}, 015002 (2019), 1805.04405.

\bibitem{Pelissetto:2000ek}
A.~Pelissetto and E.~Vicari,
\newblock Phys. Rept. {\bf 368}, 549 (2002), cond-mat/0012164.

\bibitem{Dotsenko:1979wb}
V.~Dotsenko and S.~Vergeles,
\newblock Nucl. Phys. {\bf B169}, 527 (1980).

\bibitem{Mykkanen:2012ri}
A.~Mykk{\"a}nen, M.~Panero, and K.~Rummukainen,
\newblock JHEP {\bf 1205}, 069 (2012), 1202.2762.

\bibitem{Caselle:2015tza}
M.~Caselle, A.~Nada, and M.~Panero,
\newblock JHEP {\bf 07}, 143 (2015), 1505.01106,
\newblock [Erratum: JHEP {\bf 11}, 016 (2017)].

\bibitem{Lucini:2003zr}
B.~Lucini, M.~Teper, and U.~Wenger,
\newblock JHEP {\bf 0401}, 061 (2004), hep-lat/0307017.

\bibitem{Caselle:1997dc}
M.~Caselle and M.~Hasenbusch,
\newblock Nucl. Phys. Proc. Suppl. {\bf 63}, 613 (1998), hep-lat/9709089,
\newblock [, 613 (1997)].

\bibitem{Caselle:1997hs}
M.~Caselle and M.~Hasenbusch,
\newblock J. Phys. {\bf A30}, 4963 (1997), hep-lat/9701007.

\bibitem{Hasenbusch:2010ua}
M.~Hasenbusch,
\newblock Phys. Rev. {\bf B82}, 174434 (2010), 1004.4983.

\bibitem{Caselle:2019khe}
M.~Caselle, A.~Nada, M.~Panero, and D.~Vadacchino,
\newblock JHEP {\bf 05}, 068 (2019), 1903.00491.

\bibitem{Engels:1990vr}
J.~Engels, J.~Fingberg, F.~Karsch, D.~Miller, and M.~Weber,
\newblock Phys. Lett. {\bf B252}, 625 (1990).

\bibitem{Fingberg:1992ju}
J.~Fingberg, U.~M. Heller, and F.~Karsch,
\newblock Nucl. Phys. {\bf B392}, 493 (1993), hep-lat/9208012.

\bibitem{Engels:1994xj}
J.~Engels, F.~Karsch, and K.~Redlich,
\newblock Nucl. Phys. {\bf B435}, 295 (1995), hep-lat/9408009.

\bibitem{condmat0012164}
A.~Pelissetto and E.~Vicari,
\newblock Phys. Rept. {\bf 368}, 549 (2002), cond-mat/0012164.

\bibitem{Lucini:2002ku}
B.~Lucini, M.~Teper, and U.~Wenger,
\newblock Phys. Lett. {\bf B545}, 197 (2002), hep-lat/0206029.

\bibitem{Lucini:2005vg}
B.~Lucini, M.~Teper, and U.~Wenger,
\newblock JHEP {\bf 0502}, 033 (2005), hep-lat/0502003.

\bibitem{Panero:2009tv}
M.~Panero,
\newblock Phys. Rev. Lett. {\bf 103}, 232001 (2009), 0907.3719.

\bibitem{Datta:2009jn}
S.~Datta and S.~Gupta,
\newblock Phys. Rev. {\bf D80}, 114504 (2009), 0909.5591.

\bibitem{Datta:2010sq}
S.~Datta and S.~Gupta,
\newblock Phys. Rev. {\bf D82}, 114505 (2010), 1006.0938.

\bibitem{Lucini:2012wq}
B.~Lucini, A.~Rago, and E.~Rinaldi,
\newblock Phys. Lett. {\bf B712}, 279 (2012), 1202.6684.

\bibitem{Lucini:2012gg}
B.~Lucini and M.~Panero,
\newblock Phys. Rept. {\bf 526}, 93 (2013), 1210.4997.

\bibitem{'tHooft:1976up}
G.~'t~Hooft,
\newblock Phys. Rev. Lett. {\bf 37}, 8 (1976).

\bibitem{'tHooft:1976fv}
G.~'t~Hooft,
\newblock Phys. Rev. {\bf D14}, 3432 (1976).

\bibitem{Fujikawa:1979ay}
K.~Fujikawa,
\newblock Phys. Rev. Lett. {\bf 42}, 1195 (1979).

\bibitem{Aoki:2006br}
Y.~Aoki, Z.~Fodor, S.~D. Katz, and K.~K. Szab{\'o},
\newblock Phys. Lett. {\bf B643}, 46 (2006), hep-lat/0609068.

\bibitem{Bazavov:2011nk}
A.~Bazavov {\em et~al.},
\newblock Phys. Rev. {\bf D85}, 054503 (2012), 1111.1710.

\bibitem{Alford:1998mk}
M.~G. Alford, K.~Rajagopal, and F.~Wilczek,
\newblock Nucl. Phys. {\bf B537}, 443 (1999), hep-ph/9804403.

\bibitem{Alford:2006wn}
M.~G. Alford,
\newblock PoS {\bf LAT2006}, 001 (2006), hep-lat/0610046.

\bibitem{Stephanov:2007fk}
M.~A. Stephanov,
\newblock PoS {\bf LAT2006}, 024 (2006), hep-lat/0701002.

\bibitem{Pisarski:1983ms}
R.~D. Pisarski and F.~Wilczek,
\newblock Phys. Rev. {\bf D29}, 338 (1984).

\bibitem{Pelissetto:2013hqa}
A.~Pelissetto and E.~Vicari,
\newblock Phys. Rev. {\bf D88}, 105018 (2013), 1309.5446.

\bibitem{Halasz:1998qr}
A.~M. Halasz, A.~D. Jackson, R.~E. Shrock, M.~A. Stephanov, and J.~J.~M.
  Verbaarschot,
\newblock Phys. Rev. {\bf D58}, 096007 (1998), hep-ph/9804290.

\bibitem{Berges:1998rc}
J.~Berges and K.~Rajagopal,
\newblock Nucl. Phys. {\bf B538}, 215 (1999), hep-ph/9804233.

\bibitem{Hatta:2002sj}
Y.~Hatta and T.~Ikeda,
\newblock Phys. Rev. {\bf D67}, 014028 (2003), hep-ph/0210284.

\bibitem{Antoniou:2018ico}
N.~G. Antoniou and F.~K. Diakonos,
\newblock J. Phys. {\bf G46}, 035101 (2019), 1802.05857.

\bibitem{Parotto:2018pwx}
P.~Parotto {\em et~al.},
\newblock (2018), 1805.05249.

\bibitem{Pradeep:2019ccv}
M.~S. Pradeep and M.~Stephanov,
\newblock (2019), 1905.13247.

\bibitem{Martinez:2019bsn}
M.~Martinez, T.~Sch{\"a}fer, and V.~Skokov,
\newblock (2019), 1906.11306.

\bibitem{Troyer:2004ge}
M.~Troyer and U.-J. Wiese,
\newblock Phys. Rev. Lett. {\bf 94}, 170201 (2005), cond-mat/0408370.

\bibitem{Cloet:2019wre}
J.~Arrington {\em et~al.},
\newblock {Opportunities for Nuclear Physics {\&} Quantum Information Science},
\newblock in {\em {Intersections between Nuclear Physics and Quantum
  Information Lemont, IL, USA, March 28-30, 2018}}, edited by I.~C. Clo{\"e}t
  and M.~R. Dietrich, 2019, 1903.05453.

\bibitem{Lamm:2019bik}
NuQS, H.~Lamm, S.~Lawrence, and Y.~Yamauchi,
\newblock (2019), 1903.08807.

\bibitem{Abelev:2009bw}
STAR, B.~I. Abelev {\em et~al.},
\newblock Phys. Rev. {\bf C81}, 024911 (2010), 0909.4131.

\end{thebibliography}

\end{document}